\documentclass[aps,prd,fleqn,showpacs,showkeys,twocolumn,nofootinbib,
preprintnumbers]{revtex4}
\usepackage{graphicx}
\usepackage{amssymb}
\usepackage{amsmath}
\usepackage{bm}

\begin{document}
\preprint{INHA-NTG-03/2014}
\title{In-medium modified energy-momentum tensor form factors of the
  nucleon \\ within the framework of a $\pi$-$\rho$-$\omega$ soliton model}

\author{Ju-Hyun Jung}

\affiliation{
Department of Physics, Inha University, Incheon 402-751, Republic of
Korea} 

\author{Ulugbek Yakhshiev}
\affiliation{
Department of Physics, Inha University, Incheon 402-751, Republic of
Korea} 

\author{Hyun-Chul Kim}
\affiliation{
Department of Physics, Inha University, Incheon 402-751, Republic of
Korea} 

\author{Peter Schweitzer}
\affiliation{ 
Department of Physics, University of Connecticut, Storrs,
CT 06269, U.S.A.}

\begin{abstract}
We investigate the energy-momentum tensor form factors of the nucleon
in nuclear medium, based on an in-medium modified 
$\pi$-$\rho$-$\omega$ soliton model, with medium modifications of the
mesons considered. The results allow us to establish general features
of medium modifications of the structure of nucleons bound in a
nuclear medium. 
\end{abstract}

\pacs{12.39.Dc, 21.65.Cd, 21.65.Jk}


\keywords{Energy-momentum tensor form factors, Solitonic model, Mesons
  in nuclear matter.} 

\maketitle

\section{Introduction}
The energy-momentum tensor (EMT) form factors (FFs) provide a new
aspect on the structure of the nucleon, since they contain essential 
information on how the constituents of the nucleon behave inside a  
nucleon. The EMTFFs have drawn considerable attention only very
recently, even though they were first proposed by Pagels several
decades ago~\cite{Pagels:1966zza}. The reason lies in the fact that
the natural probe to access the EMTFFs is the graviton, which is by no 
means a tractable tool to measure them experimentally. In the 
meantime the generalized parton distributions (GPDs) paved the way for
novel understanding of the inner structure of the
nucleon~\cite{Muller:1998fv, 
  Ji:1996nm,Collins:1996fb,Radyushkin:1997ki}. The EMTFFs   
are given by Mellin moments of certain GPDs and characterize how mass,
spin and internal forces are distributed inside a
nucleon~\cite{Ji:1996nm,Ji:1996ek, Polyakov:2002yz}.  
The EMTFFs are as fundamental as for instance the electromagnetic form
factors but provide different insights.

The EMTFFs of the nucleon parametrize the nucleon matrix elements of
the symmetric EMT operator as
follows~\cite{Ji:1996ek,Polyakov:2002yz}:  
\begin{widetext}
\begin{align}
\langle p^{\prime}|\hat{T}_{\mu\nu}(0)|p\rangle  \;=\;  
\bar{u}(p^{\prime},\, s')\left[M_{2}(t)\,\frac{P_{\mu}P_{\nu}}{M_{N}}
+J(t)\ 
\frac{i(P_{\mu}\sigma_{\nu\rho}+P_{\nu}\sigma_{\mu\rho})\Delta^{\rho }}{2M_{N}}
+d_{1}(t)\,\frac{\Delta_{\mu}\Delta_{\nu}-g_{\mu\nu}\Delta^{2}}{5M_{N}}\right]u(p,\, 
s)\,,
\label{Eq:EMTff} 
\end{align}
\end{widetext}
where $P=(p+p')/2$, $\Delta=(p'-p)$ and $t=\Delta^{2}$. The $M_{N}$
and $u(p,\, s)$ denote the mass and the spinor of the nucleon,
respectively. The form factor $M_{2}(t)$ is related to the
distribution of the energy density inside the nucleon. The quark and
gluon contributions to this form factor at zero-momentum transfer are
known from studies of deep-inelastic scattering and tell us that 
about a half of the momentum of a fast moving nucleon is carried by
quarks, and the other half by gluons. The form factor
$J(t)$ is related to the total angular momentum of 
quarks and gluons, and is not known experimentally. 
It is equally important to understand the form factor $d_{1}(t)$ in
Eq.~(\ref{Eq:EMTff}), since it describes how the strong forces are
distributed and stabilized in the nucleon~\cite{Polyakov:2002yz,
  Polyakov:2002wz}. In all theoretical studies so far the value of
this form factor at zero-momentum transfer, $d_1\equiv d_1(0)$, was
found to have a negative sign, and it was argued that this fact
is deeply rooted in the spontaneous breakdown of chiral
symmetry~\cite{Polyakov:1999gs,   Kivel:2000fg,Goeke:2001tz}. 
Information on EMTFFs can be obtained from studies of hard 
exclusive reactions, and in particular on $d_1(t)$ from the beam
charge asymmetry in deeply virtual Compton scattering. 

The EMTFFs of the free nucleon were studied in the bag model
\cite{Ji:1997gm} and in soliton models in Refs.~\cite{Ossmann:2004bp,
Wakamatsu:2006dy,Goeke:2007fp,Goeke:2007fq,Cebulla:2007ei}, and most 
recently in the framework of the $\pi$-$\rho$-$\omega$ solitonic
model~\cite{Jung:2013bya}. This model 
highlighted the role of $\rho$- and
$\omega$-mesons in describing the nucleon structure.
As known from the one-boson exchange potential for the  
nucleon-nucleon interaction, the vector mesons provide
short-range repulsive forces, while the pion degrees of freedom
furnish long-range attractive ones~\cite{Machleidt:1987hj}.  
The EMTFF $d_1(t)$ accommodates the ideal ground to study the interplay 
of the attractive and repulsive forces, which must exactly balance each 
other to comply with stability requirements. 
The characteristics of the vector mesons in the $\pi$-$\rho$-$\omega$ 
solitonic model turned out to be very similar to the one-boson
exchange potential: The pion provides attraction at large distances,
which is exactly balanced by repulsion at short distances due to
vector mesons. 

The nucleon is known to undergo changes in nuclear medium due to its 
environment. The EMTFFs feature essential information on how the
nucleon is modified in medium. Thus, it is very important to examine
the medium modification of the EMTFFs of the nucleon. These form
factors have been studied so far only in the framework of the
medium-modified Skyrme-model~\cite{Kim:2012ts}. In this context, it is
of great importance to extend the recent investigation in the chiral
solitonic model with explicit $\pi$, $\rho$ and $\omega$ mesonic
degrees of freedom~\cite{Jung:2013bya} to nuclear matter based on a 
medium-modified $\pi$-$\rho$-$\omega$ soliton model~\cite{Jung:2012sy}
and to study how the structure of the nucleon undergoes changes due to
the surrounding nuclear environment. The EMTFFs will directly reveal
how the changes of the $\pi$, the $\rho$, and the $\omega$ in medium
will affect the properties of the nucleon. This is our main purpose in
the present work. Moreover, the studies of nuclear medium effects  
may shed light on the EMTFFs of nuclei for which conflicting
theoretical predictions
exist~\cite{Guzey:2005ba,Liuti:2005gi,Scopetta:2009sn}.     
The first measurements of deeply virtual Compton scattering on nuclei
by HERMES \cite{Airapetian:2009bi} did not reach the level of accuracy  
required to resolve nuclear effects. But 
future experiments at Jefferson Lab 
may provide new insights into the way how the nucleon is modified in
medium. 

The present work is organized as follows: In Section II, we briefly
explain the formalism of the medium-modified $\pi$-$\rho$-$\omega$
soliton model. In Section III, we derive the expressions for the
EMTFFs within the present framework. In Section IV, we discuss the
results of the  EMTFFs. In the final Section we
summarize and draw conclusions. 
Appendix contains an alternative  general proof of the stability 
condition for the $\pi$-$\rho$-$\omega$ soliton model.

\section{General formalism}
We start from the in-medium modified effective chiral Lagrangian with
the $\pi$, $\rho$, and $\omega$ meson degrees of freedom, where
the nucleon arises as a topological soliton. 
Using the asterisk to indicate medium modified quantities,
the Lagrangian has the form
\begin{eqnarray}
\mathcal{L}^{*} & = & \mathcal{L}_{\pi}^{*}+\mathcal{L}_{V}^{*}
+\mathcal{L}_{\mathrm{kin}}^{*}+\mathcal{L}_{\mathrm{WZ}}^{*},
\label{Lag}
\end{eqnarray}
where the corresponding terms are expressed as 
\begin{eqnarray}
\mathcal{L}_{\pi}^{*} & = & \frac{f_{\pi}^{2}}{4}\,
\mbox{Tr}\left(\partial_{0}U\partial_{0}U^{\dagger}\right)
-\alpha_{p}\frac{f_{\pi}^{2}}{4}\,\mbox{Tr}
\left(\partial_{i}U\partial_{i}U^{\dagger}\right)
\cr&&
+\,\alpha_{s}\frac{f_{\pi}^{2}m_{\pi}^{2}}{2}\,\mbox{Tr}\left(U-1\right)\,,
\label{begLag}\\
\mathcal{L}_{V}^{*} & = & \frac{f_{\pi}^{2}}{2}\,
\mbox{Tr}\left[D_{\mu}\xi\cdot\xi^{\dagger}
+D_{\mu}\xi^{\dagger}\cdot\xi\right]^{2}\,,
\label{LV}\\
\mathcal{L}_{\mathrm{kin}}^{*} & = & 
-\frac{1}{2g_{V}^{2}\zeta_V}\,\mbox{Tr}\left(F_{\mu\nu}^{2}\right)\,,
\label{kin}\\
\mathcal{L}_{\mathrm{WZ}}^{*} & = & \left(\frac{N_{c}}{2}g_{\omega}
  \sqrt{\zeta_\omega}\right)\omega_{\mu}
\frac{\epsilon^{\mu\nu\alpha\beta}}{24\pi^{2}}  \cr
&&\times \,\mbox{Tr} \left\{ \left(U^{\dagger} \partial_{\nu} U
  \right) \left(U^{\dagger}\partial_{\alpha}U\right)
  \left(U^{\dagger}\partial_{\beta}U\right)\right\}. 
\label{endLag}
\end{eqnarray}
Here, the SU(2) chiral field is written as 
$U=\xi_{L}^{\dagger}\,\xi_{R}$ in unitary gauge, and the
field-strength tensor and the covariant derivative are defined,
respectively, as   
\begin{eqnarray}
F_{\mu\nu} & = & \partial_{\mu} V_{\nu} - \partial_{\nu} V_{\mu} -
i[V_{\mu},V_{\nu}]\,,
\label{Fmu}\\  
D_{\mu}\,\xi_{L(R)} & = & \partial_{\mu}\, \xi_{L(R)}-i\, V_{\mu}\,\xi_{L(R)},
\label{Dmu}
\end{eqnarray}
where the vector field $V_\mu$ includes the $\rho$-meson and 
$\omega$-meson 
fields, i.e. $\bm \rho_\mu $ and $\omega_\mu$, respectively, 
expressed as
\begin{equation}
V_\mu \;=\; \frac{g_V\sqrt{\zeta_V}}{2} (\bm \tau \cdot \bm \rho_\mu +
\omega_\mu) 
\label{Vmu}   
\end{equation}
with the Pauli matrices $\bm \tau$ in isospin space.

Note that in Eqs.~(\ref{LV}), (\ref{kin}) and (\ref{Vmu}) 
subscript $V$ generically stands for
both the $\rho$-meson and the $\omega$-meson and for compactness
we keep the generic form of those expressions. One can separate
Eqs.~(\ref{LV})  
and $(\ref{kin})$ into the $\rho$- and $\omega$-meson parts
using the definitions~(\ref{Fmu}), (\ref{Dmu}) and (\ref{Vmu}).
Then $g_V$ appearing in the $\rho$-meson part denotes $g_\rho$, 
and $g_V$ in the $\omega$-meson part designates $g_\omega$. 
Finally, $N_c=3$ is the number of colors.

Now let us discuss the parameters of the model appearing in the 
Lagrangian in Eqs.~(\ref{begLag})-(\ref{endLag}).
They can be classified into two different classes: 
(i) some of the parameters  
$f_\pi$, $m_\pi$, $g_\rho$, $g_\omega$ and $N_c$
are related to the quantities in free space while (ii) the other
parameters $\alpha_p$, $\alpha_s$ and $\zeta_V$ are
pertinent to nuclear matter properties.\footnote{$\zeta_V$ 
 denotes also a generic form for both $\zeta_\rho$ and $\zeta_\omega$ 
 which appear in the corresponding $\rho$- and $\omega$-meson parts
 of the Lagrangian.}

In free space $\alpha_p=\alpha_s=\zeta_\omega=\zeta_\rho=1$
and the free-space parameters are fixed by using either experimental or
empirical data on pions and vector 
mesons~\cite{Meissner:1986hi}.
The pion decay
constant and mass are taken to be
$f_{\pi}=93$~MeV and $m_{\pi}=135$~MeV (the neutral pion mass).   
The values of the coupling constants for the $\rho$ and $\omega$
mesons are given respectively as $g_{\rho}=5.86$ and
$g_{\omega}=5.95$. The Kawarabayashi-Suzuki--Riazuddin-Fayyazuddin
(KSRF) relation connects them to the vector meson masses,  
i.e.\ $m_{\rho}=770$~MeV and $m_{\omega}=782$~MeV, as follows  
\begin{eqnarray}
2f_{\pi}^{2}g_{\rho}^{2}=m_{\rho}^{2}\,,\qquad 2f_{\pi}^{2}g_{\omega}^{2}
=m_{\omega}^{2}\,.
\end{eqnarray}

In general, the parameters $\alpha_p$,
$\alpha_s$, and $\zeta_V$ stand for the medium functionals which are the
essential quantities in the present work.  
They depend on the nuclear matter density $\rho$ and are defined as
\begin{eqnarray}
\alpha_{p}(\rho)  &=&1-\frac{4\pi c_{0}\rho/\eta}{1+g_{0}'4\pi
  c_{0}\rho)/\eta}\,,\cr 
\alpha_{s} (\rho) &=&1-{4\pi\eta b_{0}\rho}{m_{\pi}^{-2}}\,,\cr
\zeta_V(\rho) &=& \exp\left\{ -\frac{\gamma_{{\rm num}}\rho}
  {1+\gamma_{{\rm den}}\rho}\right\}  \,.
\label{medfunc}
\end{eqnarray}
They provide crucial information on how the nuclear-matter environment
influences properties of the single soliton~\cite{Jung:2012sy}. 
The $\eta$ is a kinematic factor defined as
$\eta=1+m_{\pi}/m_{N}\simeq1.14$. The values of the empirical parameters 
$b_{0}=-0.024\, m_{\pi}^{-1}$ and $c_{0}=0.09\, m_{\pi}^{-3}$ are taken from the
analysis of pionic atoms and the data on low-energy pion-nucleus
scattering. The $g_{0}'=0.7$ denotes the Lorentz-Lorenz factor that takes
into account the short-range correlations~\cite{Ericsonbook}. 

The additional parameters  $\gamma_{\rm num}$ and $\gamma_{\rm den}$ 
are introduced phenomenologically to reproduce the saturation 
point at normal nuclear matter. Two different techniques have 
been discussed in literature~\cite{Jung:2012sy}, in order to introduce
nuclear modifications in the present soliton approach, 
and in this work we will explore both models.

{\it Model I}: Here one neglects the small mass difference of the 
$\rho$- and $\omega$-mesons in free space ($m_\omega=m_\rho=770$~MeV, 
$g_\omega=g_\rho=5.86$) and assumes that the KSRF relation still 
holds in nuclear matter 
\begin{eqnarray}
2f_{\pi}^{2}g_{\rho}^{2}\zeta_\rho&=& m_\rho^{*2} \;=\;
m_\omega^{*2},\quad \zeta_\rho=\zeta_\omega\neq 1. 
\label{eq:KSRF1}
\end{eqnarray}

{\it Model II}: Here we remove the degeneracy of the vector meson
masses in free space ($m_\rho\neq m_\omega=782$~MeV, $g_\rho\neq 
g_\omega=5.95$), and instead of Eq.~(\ref{eq:KSRF1}) assume that
the KSRF relation is valid only for the $\rho$ meson, with the
$\omega$ meson kept as in free space: 
\begin{eqnarray}
2f_{\pi}^{2}g_{\rho}^{2}\zeta_\rho&=& m_\rho^{*2} \;\neq\;
m_\omega^{*2}, \quad \zeta_\rho \neq 1, \;\;\;\; \zeta_\omega =1.
\label{eq:KSRF2}
\end{eqnarray}

The two different techniques to implement nuclear 
modifications within the current approach reflect the
possibility that the $\rho$- and $\omega$-degrees of freedom 
could respond differently to a nuclear medium 
environment~\cite{Naruki:2005kd,Wood:2008ee}. 
The effects of the $\omega$-mesons are mainly limited to the inner  
core of the nucleon. Therefore, the two variants describe the 
situation that the inner core of the nucleon is more (Model I) 
or less (Model II) affected by medium effects. The latter is
a plausible scenario, at least around the normal nuclear 
matter density.

In practice, these two models yield comparable results 
in many respects. A notable (and in our context important) 
difference though, is the description of the incompressibility 
of symmetric nuclear matter: 
Model I yields a smaller value of the incompressibility, while 
Model II produces a larger one. It means that  
Model II gives a stiffer nuclear binding energy and agrees better  
with the data~\cite{Jung:2012sy}. 
In both models the values of $\gamma_{\mathrm{num}}$
and $\gamma_{\mathrm{den}}$ are fitted to reproduce the coefficient of
the volume term in the empirical mass formula $a_V\approx 26$~MeV.
Although this is larger than the experimental value 
$a_V^{\rm exp}\approx 16$~MeV, the relative change 
of the in-medium nucleon mass is reproduced correctly. 
(See Eq.~(12) in Ref.~\cite{Jung:2012sy} and the corresponding
explanation.) In Model I we have $\gamma_{{\rm
    num}}=2.390\,m_{\pi}^{-3}$ and $\gamma_{{\rm den}}=1.172\,
m_{\pi}^{-3}$, whereas in Model II we employ $\gamma_{{\rm
    num}}=1.970\, m_{\pi}^{-3}$ and $\gamma_{{\rm den}}=0.841\,
m_{\pi}^{-3}$. For further details on these two models in relation  
to nuclear matter properties, we refer to Ref.~\cite{Jung:2012sy}.

Since we are interested in \textit{homogeneous} and \textit{symmetric}
nuclear matter, the nuclear matter density can be regarded as a
constant and the \textit{spherically symmetric} hedgehog Ans\"atze can
be employed. (In this situation the medium functionals $\alpha_p$,
$\alpha_s$, $\zeta_V$ become ordinary functions of the 
nuclear matter density $\rho$.) 
With the notation $ \hat{\bm{n}} = {\bm{x}}/r$
and $r=|{\bm{x}}|$, the chiral soliton and the vector fields
can be expressed in terms of the radial profile functions $F(r)$,
$G(r)$, and $\omega(r)$: 
\begin{eqnarray}
U & = & \exp\left\{i\bm{\tau} \cdot \hat{\bm{n}} F(r)\right\}
\,,\cr 
\rho_{\mu}^{a}&=& \frac{\varepsilon_{0\mu
    ka}\hat{n}_{k}}{g_\rho\sqrt{\zeta_\rho}r}\,G(r)\,,\cr
\omega_{\mu}&=& \omega(r) \delta_{\mu0}. \label{Eq:14}
\end{eqnarray}
Note that the presence of the factor $(g_\rho\sqrt{\zeta_\rho})^{-1}$ in
$\rho_i^a$ is essential to keep the same boundary conditions for the
medium-modified profile function $G(r)$ as in free space,
i.e. $G(0)=-2$. 

Utilizing the hedgehog Ans\"atze, we are able to derive the static
energy functional from the Lagrangian, which is identified as the
classical soliton mass:  
\begin{widetext}
\begin{eqnarray}
\label{Eq:XXXX}
M_{\mathrm{sol}}^* & = &4\pi \int_0^\infty {\rm d}r\, r^2\,
\left\{\alpha_{p}f_{\pi}^{2}\left(\frac{\sin^{2}F}{r^{2}}+
\frac{F'^{2}}{2}\right)\right.
+\,\alpha_{s}f_{\pi}^{2}m_{\pi}^{2}\left(1-\cos F\right)
+\frac{2f_{\pi}^{2}}{r^{2}}\left(1-\cos F+G\right)^{2}
-\zeta_\omega g_\omega^{2}f_{\pi}^{2}\omega^{2}
\cr&& 
+\frac{1}{g_\rho^{2}\zeta_\rho
  r^{2}}\left(G'^{2}+\frac{G^{2}\left(G+2\right)^{2}}{2r^{2}}\right) 
-\frac{1}{2}\omega'^{2}
\left.+\left(\frac{3}{2}g_\omega
\sqrt{\zeta_\omega}\right)\frac{1}{2\pi^{2}r^{2}}
\omega\sin^{2}F\, F'\right\},
\end{eqnarray}
\end{widetext}
where $f'= \partial f/\partial r$, generically.  

The next step is to minimize the classical soliton mass. This is
done by solving the equations of motion for each meson field, 
which are derived as 
\begin{eqnarray}
F''&\!=\!&-\frac{2}{r}F'+\frac{1}{\alpha_{p}r^{2}}
\Big(4\left(G+1\right ) \sin F-(2-\alpha_{p})\sin2F\Big) \cr
&& + \frac{\alpha_{s}m_{\pi}^{2}}{\alpha_{p}}\sin F
-\frac{3g_\omega\sqrt{\zeta_\omega}}{4\pi^{2}\alpha_{p}f_{\pi}^{2}}\frac{\sin^{2}
  F\omega'}{r^{2}}\label{eq:F}\,,\\ 
G''&\!=\!&2f_{\pi}^{2}g_\rho^{2}
\zeta_\rho \left(G+1\!-\!\cos{F}\right)
+ \frac{\left(G+2\right)\left(G+1\right)G}{r^{2}}\,, \qquad
\label{eq:G}\\
\omega''&\!=\!& -\frac{2}{r}\omega' + 2f_{\pi}^{2}g_\omega^{2}
\zeta_\omega\omega
- \frac{3g_\omega\sqrt{\zeta_\omega}}
{4\pi^{2}r^{2}} F'\sin^{2}F
\label{eq:w}
\end{eqnarray}
with the corresponding boundary conditions
\begin{eqnarray}
&&F(0)=\pi, \quad G(0)=-2,\quad \omega'(0)=0,\cr
&&F(\infty)= G(\infty)= \omega(\infty)=0.
\end{eqnarray}

Having quantized the soliton collectively, we obtain 
\begin{eqnarray}
U(\bm{r},\,t) &=& A(t)U(\bm{r})A^+(t)\,,\cr 
\omega_i(\vec r,t) &=& \frac{\Phi(r)}{r}\, \left(\bm{K}\times
  \frac{\bm{r}}r\right)_i, \cr 
\bm{\tau}\cdot\bm{\rho_0} (\bm{r},\,t) &=& \frac{2}{g_\rho} A(t)
\bm{\tau}\cdot\left[\bm{K} \xi_1(r) \right.\cr
&& \left. \;\;\;\; + \; \hat{\bm n}
  \left(\bm{K} \cdot \hat{\bm n} \right) \xi_2(r)\right] A^+(t),\cr
\bm{\tau}\cdot\bm{\rho}_i(\bm{r},\,t) &=& A(t) \bm{\tau} \cdot
\bm{\rho}_i(\bm{r}) A^+(t),
\end{eqnarray}
where $2\bm{K}$ denotes the angular velocity of the soliton with
the relation $i\bm{\tau}\cdot \bm{K} = A^+\dot{A}$. This leads to the
time-dependent collective Hamiltonian 
\begin{equation}
H(t) = M_{\mathrm{sol}}^* +\lambda^* {\rm Tr}(\dot A\dot A^+),   
\end{equation}
where $\lambda^*$ denotes the moment of inertia for the rotating
soliton  
\begin{widetext}
\begin{eqnarray}
\lambda^* & = & 4\pi\int_0^\infty\!\!
dr\left\{\frac{2}{3}f_{\pi}^{2}r^{2}\left(\sin^{2}F+ 
8\sin^{4}\frac{F}{2}-8\xi_{1}\sin^{2}\frac{F}{2}+3\xi_{1}^{2}
+2\xi_{1}\xi_{2}+\xi_{2}^{2}\right)\right. \cr
 &  & \;\;\;+\; \frac{1}{3g_\rho^{2}\zeta_\rho}\Big(
4G^{2}\left(\xi_{1}^{2}+\xi_{1}\xi_{2}-2\xi_{1}-\xi_{2}
  +1\right)+2\left(G^{2}+2G+2\right)\xi_{2}^{2} 
+r^{2}\left(3\xi_{1}^{\prime2}+\xi_{2}^{\prime2}+2\xi_{1}'\xi_{2}'
\right)\Big) \cr 
 &  &  \qquad\qquad\left.-\frac{1}{6}
   \left(\Phi'^{2}+\frac{2\Phi^{2}}{r^{2}}+  
2f_{\pi}^2g_\omega^2{\zeta_\omega}\Phi^{2}\right)
+g_\omega\sqrt{\zeta_\omega} \frac{\Phi
F'}{2\pi^{2}}\sin^{2}F\right\}\,. 
\label{eq:moi}
\end{eqnarray}
\end{widetext}
In the large $N_{c}$ limit, we are able to minimize the moment of
inertia, so that we derive the coupled nonlinear differential
equations for the next-order profile functions $\xi_{1}$, $\xi_{2}$ and
$\Phi$ 
\begin{eqnarray}
\xi_{1}'' & = & 2f_{\pi}^{2}g_\rho^{2}\zeta_\rho
\left(\cos F+\xi_{1}-1\right)\nonumber \\
 &  &
 \quad+\frac{G^{2}\left(\xi_{1}-1\right)
   +2(G+1)\xi_{2}}{r^{2}}-\frac{2\xi_{1}'}{r},\\  
\xi_{2}'' & = & 2f_{\pi}^{2}g_\rho^{2}\zeta_\rho 
\left(-\cos F+\xi_{2}+1\right)
-\frac{2\xi_{2}'}{r}\nonumber \\
 &  & \quad+\frac{G^{2}\left(\xi_{1}-1\right)
+2\left[\left(G+3\right)G+3\right]\xi_{2}}{r^{2}},\quad\\
\Phi'' & = & 2f_{\pi}^{2}g_\omega^{2}
\zeta_\omega \Phi
-\frac{3g_\omega\sqrt{\zeta_\omega}
F'\sin^{2}F}{2\pi^{2}}+\frac{2\Phi}{r^{2}}
\end{eqnarray}
with the boundary conditions defined as 
\begin{eqnarray}
&&\xi_{1}'(0)= \xi_2'(0) =\Phi(0)=0,\cr
&&\xi_{1}(\infty)=\xi_2(\infty)=\Phi(\infty)=0.
\end{eqnarray}
The boundary conditions for $\xi_1$ and $\xi_2$ are constrained to
satisfy $2\xi_1(0)+\xi_2(0)=2$ that remains unchanged in nuclear
matter. The other details can be found in Ref.~\cite{Jung:2012sy}. 

\section{Energy-momentum tensor form factors}
Using the definition of the canonical EMT operator 
\begin{equation}
T^{\mu\nu*} \;=\; \frac{\partial \mathcal{L^*}}{\partial (\partial
  \phi_a)}\partial^\nu \phi_a - g^{\mu\nu} \mathcal{L^*},
\end{equation}
where $\phi_a$ are the relevant meson degrees of freedom and
$g^{\mu\nu}=\mathrm{diag}(1,\,-1,\,-1\,,-1)$ is the metric 
tensor in Minkowski space, and the Lagrangian as defined 
in Eq.~(\ref{Lag}), we can derive the expressions for the 
densities of the EMT. The resulting EMT is symmetric
and the expressions for its components are given as 
\begin{widetext}
\begin{eqnarray}
T^{00*}\left(r\right) & = & \alpha_{p}\frac{f_{\pi}^{2}}{2}
\left(2\frac{\sin^{2}F}{r^{2}}+F'^{2}\right) 
+\alpha_{s}f_{\pi}^{2}m_{\pi}^{2}\left(1-\cos F\right)
+\frac{2f_{\pi}^{2}}{r^{2}}\left(1-\cos F+G\right)^{2} - 
f_{\pi}^{2}g_\omega^{2}\zeta_\omega\omega^{2}\cr 
&& + \frac{1}{g_\rho^{2}\zeta_\rho r^{2}}\left(G'^{2}+\frac{G^{2}
  \left(G+2\right)^{2}}{2r^{2}}\right)-\frac{1}{2}\omega'^{2} 
+\left(\frac{3}{2}g_\omega\sqrt{\zeta_\omega}\right)
\frac{1}{2\pi^{2}r^{2}}\omega\sin^{2}F\,F'\,,
\end{eqnarray}
\end{widetext}
\begin{eqnarray}
&& \hspace{-5mm}
T^{0i*}\left(\bm{r},\bm{s}\right) = 
\frac{e^{ilm}r^{l}s^{m}}{\left(\bm{s}\times
    \bm{r}\right)^{2}}\rho_{J}^{*}\left(r\right)\,,\\ 
\label{Eq:30}
&& \hspace{-5mm}
T^{ij*}\left(r\right)  =
s^{*}\left(r\right)\left(\frac{r^{i}r^{j}}{r^{2}} 
-\frac{1}{3}\delta^{ij}\right)+p^{*}\left(r\right)\delta^{ij}\,,\;\;\;
\end{eqnarray}
where $T_{00}(r)$ is called the energy density, which provides
information on how the mass is distributed inside a nucleon. The
vector $\bm s$ represents the direction of the quantization axis for
the spin and coincides with the space part of the polarization vector
of the nucleon in the rest frame. The densities $\rho_J^*(r)$,
$p^*(r)$ and $s^*(r)$ denote respectively the angular momentum, 
pressure and shear-force densities in nuclear matter, which are
derived as  
\begin{widetext}
\begin{eqnarray}
\rho_{J}^{*}\left(r\right) & = & \frac{f_{\pi}^{2}}{3\lambda^*}
\left(\sin^{2}F+8\sin^{4}\frac{F}{2}-4\sin^{2}\frac{F}{2}\xi_{1}\right)
+\frac{1}{3g_\rho^{2}\zeta_\rho
r^{2}\lambda^*}\left[\left(2-2\xi_{1}-\xi_{2}\right)
  G^{2}\right]
+\frac{g_\omega\sqrt{\zeta_\omega}}{8\pi^{2}\lambda^*}\Phi\sin^{2}FF'\,,
\label{eq:rhoj}\\
p^{*}\left(r\right)& = & -\frac{1}{6}\alpha_{p}f_{\pi}^{2}
\left(F'^{2}+2\frac{\sin^{2}F}{r^{2}}\right) 
-\alpha_{s}f_{\pi}^{2}m_{\pi}^{2}\left(1-\cos F\right)
-\frac{2}{3r^{2}}f_{\pi}^{2}\left(1-\cos F+G\right)^{2} +
f_{\pi}^{2}g_\omega^{2}\zeta_\omega\omega^{2} 
+\frac{1}{6}\omega'^{2}\nonumber \label{Eq:32}\\
 &  & +\frac{1}{3g_\rho^{2}\zeta_\rho r^{2} }\left(G'^{2}+\frac{G^{2}
   \left(G+2\right)^{2}}{2r^{2}}\right) \,,\\ 
s^{*}\left(r\right) & = & \alpha_{p}f_{\pi}^{2}
\left(F'^{2}-\frac{\sin^{2}F}{r^{2}}\right) 
-\frac{2f_{\pi}^{2}}{r^{2}}\left(1-\cos F+G\right)^{2}
+\frac{1}{g_\rho^2\zeta_\rho r^{2}}\left(
  G'^{2}-\frac{G^{2}\left(G+2\right)^{2}}{r^2}\right)
-\omega'^{2}\,. \label{Eq:33} 
\end{eqnarray}
\end{widetext}

We will follow Refs.~\cite{Cebulla:2007ei, Kim:2012ts} to compute the
EMTFFs in a consistent way, and consider for each quantity only
the respective leading contribution in the large $N_c$ limit. Thus,
the EMTFFs are derived as   
\begin{eqnarray}
M_2^*(t)-\frac{t\,d_1^*(t)}{5M_N^{*2}}
        &\!=\!& \frac{1}{M_N^*}\int\mathrm{d}^3
        \bm{r}\;T_{00}^*(r)\;j_0(r\sqrt{-t})\,,
        \label{Eq:M2-d1-model-comp}\\
        d_1^*(t)
        &\!=\!& \frac{15 M_N^*}{2}\int\mathrm{d}^3 \bm{r} \;p^*(r)
        \;\frac{j_0(r\sqrt{-t})}{t} \,,
        \label{Eq:d1-model-comp}\qquad\\
        J^*(t)
        &\!=\!& 3
    \int\mathrm{d}^3\bm{r}\;\rho_J^*(r)\;\frac{j_1(r\sqrt{-t})}{r\sqrt{-t}}\;,
    \label{Eq:J-model-comp}
\end{eqnarray}
where $j_0(z)$ and $j_1(z)$ represent the spherical Bessel functions 
of order 0 and 1, respectively. 

Some of the EMT densities are required to satisfy the following
constraints 
\begin{eqnarray}
\label{eq:mass_con} 
\frac{1}{M_N^*} \int\mathrm{d}^3 \bm{r}\; T_{00}^*(r)&=&M_2^*(0)\; =
\; 1\,,\\ 
\label{eq:norm}\int\mathrm{d}^3\bm{r}\;\rho_{J}^*(r)&=&J^*(0) \; = \;
\frac12,\\ 
\label{Eq:39} \int\mathrm{d}^3\bm{r}\;p^*(r)  &=& 0. 
\end{eqnarray}
Equation~(\ref{eq:mass_con}) means that the volume 
integral of the energy density reproduces the nucleon mass. The second  
constraint given in Eq.~(\ref{eq:norm}) indicates that the nucleon has
spin $\frac12$, and the third one (\ref{Eq:39}) 
is the necessary condition for stability~\cite{von-Laue} of the
nucleon and is also known as the von Laue condition 
\cite{BialynickiBirula:1993ce}.

The nuclear medium modifies the dynamics inside the nucleon as compared 
to the free space case, which may result in quantitatively distinct
features. However, the general soliton structure in medium is kept to
be the same as in free space. The proofs of the relations
(\ref{eq:mass_con}--\ref{Eq:39}) for a nucleon in medium, can 
therefore be step by step carried over from corresponding proofs given
for a free nucleon in Ref.~\cite{Jung:2013bya}, and we do not repeat
them here.  In Appendix~\ref{App:A} we give an alternative proof of
the von Laue condition (\ref{Eq:39}).

\section{Results and discussions}
\begin{table*}[ht]
\begin{ruledtabular}
\begin{tabular}{ccccccccc}
&
{$T_{00}(0)$}& 
{$\left\langle r_{00}^{2}\right\rangle $} & 
{$\left\langle r_{J}^{2}\right\rangle $} & 
{$p(0)$} & 
{$r_{0}$} & 
{$d_{1}(0)$} &
$\langle r_F^2\rangle$\\
 &
{$\left[\mbox{GeV}/\mbox{fm}^{3}\right]$} & 
{$\left[\mbox{fm}^{2}\right]$} & 
{$\left[\mbox{fm}^{2}\right]$} & 
{$\left[\mbox{GeV}/\mbox{fm}^{3}\right]$} & 
{$\left[\mbox{fm}\right]$} & &  
{$\left[\mbox{fm}^{2}\right]$}
\smallskip\\
\hline
Present work (Model I) 
&3.56 & 0.78 & 0.74 & 0.58 & 0.55 &  $-$5.03 & 1.00 \\  
Present work (Model II) 
&3.51 & 0.79 & 0.74 & 0.59 & 0.55 &  $-$5.13 & 1.01 \\ 
Skyrme model~\cite{Kim:2012ts}
&
1.45  & 0.68 & 1.09 & 0.26 & 0.71 & $-$3.54  & 1.10 
\end{tabular}
\end{ruledtabular}
\caption{The quantities relevant to the nucleon EMT densities and
  their form factors for the free space nucleons: 
  $T_{00}\left(0\right)$ is the energy density in
  the center of the nucleon; $\langle r_{00}^{2}\rangle$ is the
  mean-squared radius of the energy density;  $\langle
  r_{J}^{2}\rangle$ represent that of the angular momentum
  distributions; $p(0)$ denotes the pressure value at 
  the origin;  $r_{0}$ designates the node of the pressure
  distribution such that $p(r_0)=0$; $d_{1}(0)$ corresponds to the 
$d_{1}\left(t\right)$ form factors at zero momentum transfer;
and $\langle r_F^2\rangle$ is defined below in Eq.~(\ref{rF}). 
{\label{tab:1}}}
\end{table*}

In this Section, we present and discuss the numerical results.
In Table~\ref{tab:1}, we list the results of the quantities relevant
to the EMT densities and FFs in free space. One can see that both the 
results from Model~I and Model~II are very similar to each
other, which indicates that the differences of the masses and the
coupling constants between the $\rho$ meson and the $\omega$ meson
parameters are not crucial in free space.
The results from Model I were reported in Ref.~\cite{Jung:2013bya}. 

\begin{table*}[bt]
\begin{ruledtabular}
\begin{tabular}{cccccccc}
&{$T_{00}^{*}(0)/T_{00}(0)$}& 
{$\left\langle r_{00}^{2}\right\rangle^*/
\left\langle r_{00}^{2}\right\rangle$} & 
{$\left\langle r_{J}^{2}\right\rangle^*/
\left\langle r_{J}^{2}\right\rangle $} & 
{$p^*(0)/p(0)$} & 
{$r_{0}^*/r_0$} & 
{$d_{1}^*(0)/d_1(0)$} & 
$\langle r_F^2\rangle^\ast$/$\langle r_F^2\rangle$\\
\smallskip\\
\hline
Present work (Model I) & 0.61 & 1.36 & 1.04 & 0.59 & 1.18 & 1.11 & 1.31\\
Present work (Model II)& 0.52 & 1.46 & 1.09 & 0.59 & 1.22 & 1.33 & 1.44\\
Skyrme 
model~\cite{Kim:2012ts}& 0.49 & 1.40 & 1.24 & 0.50 & 1.27 & 1.37 & 1.41\\
\end{tabular}
\end{ruledtabular}
\caption{The ratio of the quantities at normal nuclear matter 
density $\rho=\rho_0$  
to those in free space presented 
in Table~\ref{tab:1}. For comparison the results
from the in-medium modified Skyrme model~\cite{Kim:2012ts} 
are also presented. 
{\label{tab:2}}}
\end{table*}

The relative in-medium changes in the EMT densities and the EMTFFs 
are presented in Table~\ref{tab:2}. 
For comparison we include the results from the Skyrme
model without (in Table~\ref{tab:1})  
and with (in Table~\ref{tab:2}) medium modifications~\cite{Kim:2012ts}.
Depending on the quantity, the absolute numbers from the Skyrme
model of Ref.~\cite{Kim:2012ts} and the $\pi$-$\rho$-$\omega$ soliton 
model studied here differ significantly, see Table~\ref{tab:1}. 
For instance, the energy density and the pressure in the center of the  
nucleon are lower in the Skyrme model~\cite{Kim:2012ts} than
those of the present work. In large part, these differences can be
attributed to the smaller soliton mass $M_{\rm sol}=881\,{\rm MeV}$ in 
the Skyrme model~\cite{Kim:2012ts} as compared to $1473\,{\rm MeV}$ in
the present work. However, the aim of this work is not to compare the 
Skyrme and $\pi$-$\rho$-$\omega$ soliton models {\it per
  se}.~\footnote{That would require tuning model parameters to have
  more comparable soliton descriptions. For more discussion of Skyrme
  and $\pi$-$\rho$-$\omega$ soliton models for a free nucleon, from
  the point of view of EMT properties, see Ref.~\cite{Jung:2013bya}.} 
Rather this work is focused on studying nuclear medium effects. 
It is therefore a remarkable observation that,
in spite of numerically very different {\it absolute} descriptions,
in Table~\ref{tab:1}, the {\it relative impact} of a nuclear
environment on the properties of a single nucleon is qualitatively 
similar in the $\pi$-$\rho$-$\omega$ soliton- and Skyrme-model,
see Table~\ref{tab:2}. 

The energy density in the center of the nucleon, $T_{00}^{*}(0)$, is
rather sensitive to the change of the nuclear matter density. The
value of $T_{00}^{*}$ from Model I is reduced by around $40\,\%$ and
that from Model II decreases by almost $50\,\%$. This is in line with
the observations made in the Skyrme model~\cite{Kim:2012ts}, where
$T_{00}^{*}$ also decreased by about $50\,\%$. It indicates to some
degree the well known fact that the nucleon mass tends to decrease in
nuclear matter.  

The energy mean-squared radius $\langle r_{00}^2 \rangle^*$ 
is defined as 
\begin{equation}
\langle r_{00}^2 \rangle^* \;=\; \frac{\int d^3 \bm r \, r^2
  \,T_{00}^* (r)}{\int d^3 \bm r \,  \,T_{00}^* (r)}   \,.
\end{equation}
As shown in Table~\ref{tab:1}, the value of $\langle r_{00}^2
\rangle^*$ increases from $0.78\,(0.79)\,\mathrm{fm}^2$ to
$1.06\,(1.15) \mathrm{fm}^2$ in the case of Model I (Model II) when
the density changes from zero to the normal nuclear matter density. 
The rate of the change in $\langle r_{00}^2 \rangle^*$ is almost the
same as in the case of the Skyrme model (see Table~\ref{tab:2}). 
It implies that the size of
the nucleon in medium generally increases. We will discuss the
physical implications of other quantities along with the EMTFFs.  

\begin{figure*}
\includegraphics[scale=0.4]{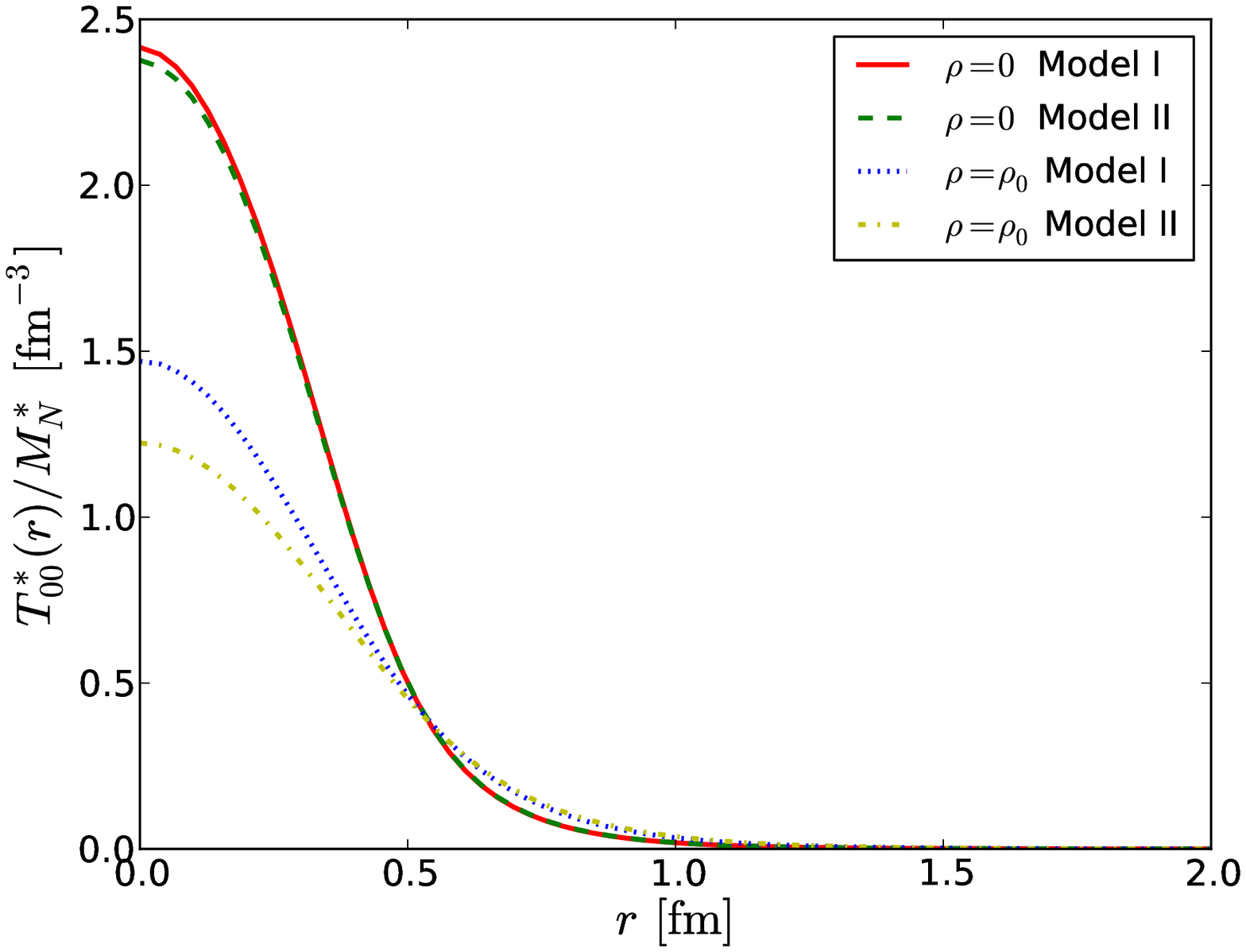}\hskip 1cm
\includegraphics[scale=0.4]{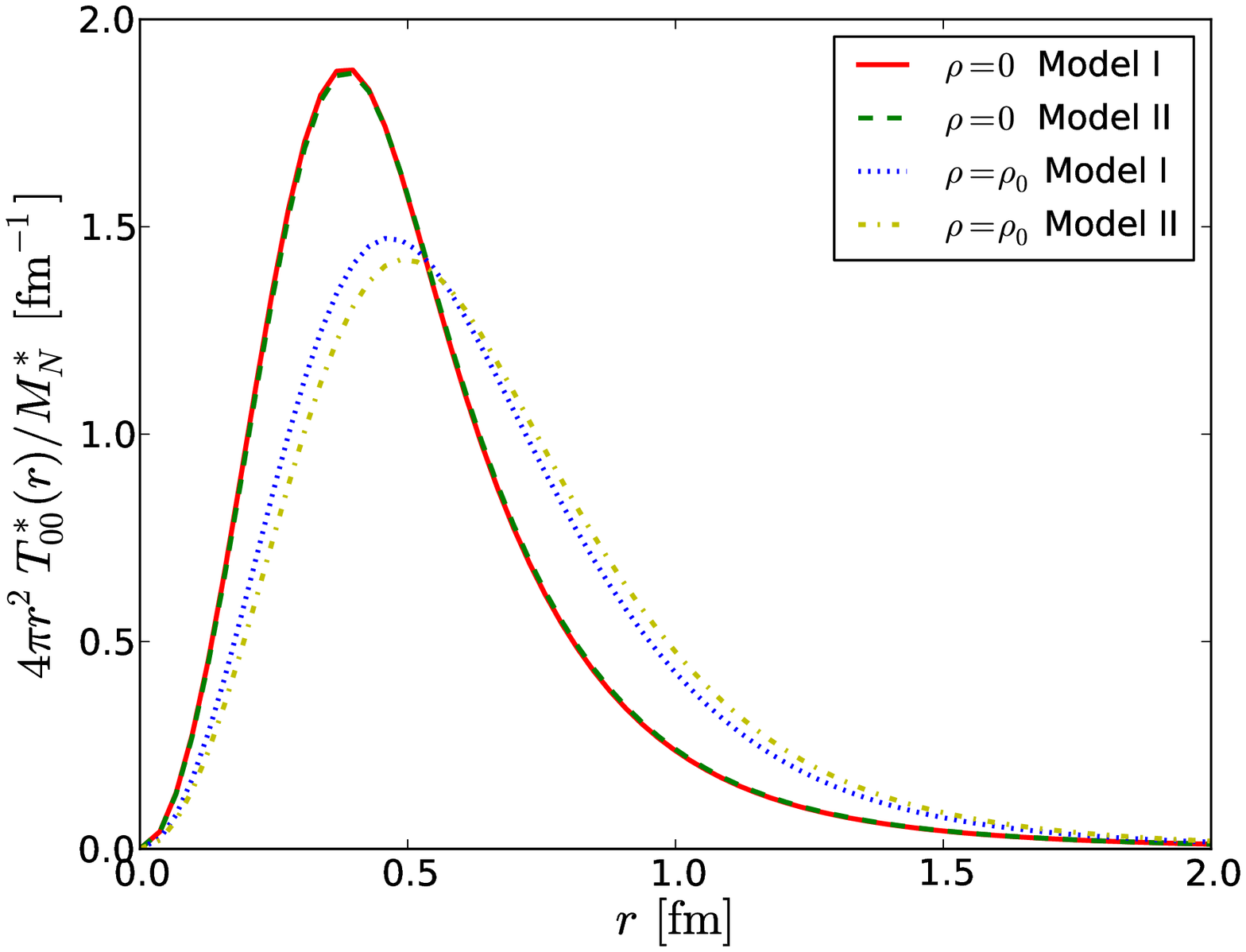} 
\caption{The energy density in the nucleon normalized by the nucleon
  mass, $T_{00}^{*}(r)/M_{N}^{*}$ in the left panel and the $4\pi
  r^{2}T_{00}^{*}(r)/M_{N}^{*}$ in the right panel, as functions of
  radial distance $r$. The solid and dashed curves depict the
  densities respectively from Model I and Model II in free space. The
  dotted and dot-dashed ones represent respectively those  
  from Model I and Model II in nuclear matter. }
\label{fig1}
\end{figure*}
In the left panel of Fig.~\ref{fig1} we depict the results 
of the energy density 
normalized by the nucleon mass as functions of the radial distance
$r$. While there is almost no difference between Model I and Model II
in free space, we find that $T_{00}^{*}(r)$ in nuclear matter
undergoes a larger change in Model II than in Model I.  This can be
traced back to the 
fact that the $\omega$ meson in Model II is assumed to remain intact 
in nuclear matter, while in Model I both the $\rho$ and $\omega$
meson masses undergo medium modifications.  
In general, the energy density is broadened and its
maximum is reduced in nuclear medium. The broadening of the energy 
density can be more clearly seen in the density weighted by  
the factor $4\pi r^2$ as shown in the right panel of 
Fig.~\ref{fig1}, which results in the above-discussed 
increase of the energy mean-squared radius $\langle r_{00}^2\rangle^*$.

\begin{figure}
\includegraphics[scale=0.4]{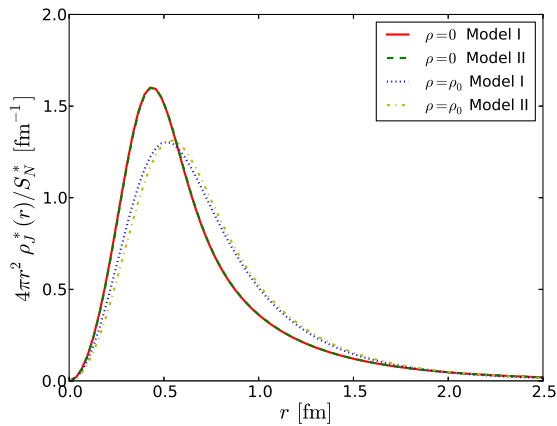}
\caption{The angular momentum density $\rho_{J}^{*}(r)$ normalized
by the nucleon spin $S_{N}^{*}=1/2$ as a function of radial distance
$r$. The solid and dashed curves depict the
  densities respectively from Model I and Model II in free space. The
  dotted and dot-dashed ones represent respectively those  
  from Model I and Model II in nuclear matter.} 
\label{fig2}
\end{figure}

Figure~\ref{fig2} draws the spin densities of the nucleon as
a function of $r$, normalized by its spin. 
As in the case of the energy densities, 
the spin density of the nucleon is also broadened in
nuclear matter and its maximum is lessened. Note that the 
integration of the spin density over the space gives the spin of
the nucleon, as defined in Eq.~(\ref{eq:norm}). 
This means that the decrease of the magnitude of the
spin density in nuclear matter is compensated by the broadening 
of the density in such a way that the nucleon spin turns out be always
one half. The spin mean-squared radius
increases slightly in nuclear matter as shown in Table~\ref{tab:2},
which is interesting because the $\langle r_{00}^2\rangle^*$ exhibits
the increment in nuclear matter by about $40\,\%$ whereas the $\langle
r_{J}^2\rangle^*$ increases by less than $10\,\%$. 

Technically, the small changes in the angular momentum distribution 
can be understood from Eq.~(\ref{eq:rhoj}). One can see that the 
distribution corresponding to the external part of the soliton has 
no medium factor.
This is due to the fact that the time-dependent part of the Lagrangian 
${\cal L}_\pi^*$ is not modified in nuclear matter. As a result, 
the contribution from the outer shell of the rotating soliton remains 
more or less the same as in free space. Although the inner
part of the soliton has explicit medium factors, clearly
they will not lead to larger changes.\footnote{The situation becomes 
different when one considers the effects of explicit isospin symmetry
breaking~\cite{Yakhshiev:2013eya} due to the explicit medium
modification of the time-dependent part of the ${\cal L}_\pi^*$.}  
The medium-modified Skyrme model exhibits a similar 
tendency~\cite{Kim:2012ts}, though the relative increase of
$\langle r_J^2\rangle^*$ is more substantial in that model, 
see Table~\ref{tab:2}.

\begin{figure}
\includegraphics[scale=0.4]{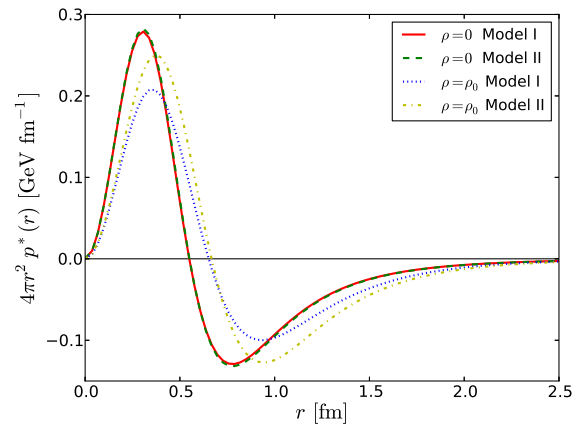}
\caption{The pressure density $p^{*}(r)$ as a function of radial
  distance $r$. The solid and dashed curves depict the
  densities respectively from Model I and Model II in free space. The
  dotted and dot-dashed ones represent respectively those  
  from Model I and Model II in nuclear matter. }
\label{fig3}
\end{figure}

Figure~\ref{fig3} depicts the pressure densities in a free
nucleon and in a nucleon in nuclear matter. 
In fact, the pressure density is the
most interesting quantity in understanding the modification of the
nucleon in nuclear matter, because it reveals vividly the
internal dynamics in the nucleon. As shown in Fig.~\ref{fig3}, 
the pressure density is positive in the inner region, and 
negative in the outer region. This must be so on general grounds. 
Positive pressure in the inner region signals repulsive forces, 
negative pressure in the outer region means attractive forces. 
Repulsive forces in the inner and attractive forces in the outer 
parts must balance each other exactly according to Eq.~(\ref{Eq:39}). 

As in the case of the energy and spin densities, 
we observe that the pressure density is also broadened in 
nuclear matter and its modulus is reduced. 
In Table~\ref{tab:2}, we list the values of the
pressure density at $r=0$, i.e. $p^*(0)$, and those of $r_0^*$, which 
designate the position in which the sign of the pressure density
changes. The simultaneous overall decrease of the modulus of $p(r)$, 
and the broadening of the pressure density in nuclear matter
occur in such a way that the results of the pressure density 
in medium also comply with the stability condition~(\ref{Eq:39}).

It is interesting to compare the pressure distributions
in Model I and Model II.
One can see that the absolute value of the pressure 
in Model II is larger in comparison with that in Model I.
In the present approach, the nucleon in nuclear environment 
is a stable object by itself. No ``external forces'' due to 
the nuclear medium are required to stabilize it. However, 
nuclear matter has an impact on the nucleon. In this sense, 
it is a physically appealing observation that the stiffer nuclear 
matter and the higher incompressibility in Model II \cite{Jung:2012sy}  
cause the nucleon to be subject to stronger internal forces, for
which the magnitude of $p(r)$ is a measure.
The stronger repelling inner forces in Model II make moreover
the nucleon swell more strongly in nuclear matter. We have seen 
above the consequence of this: the relative decrease in energy 
density in the nucleon core is more pronounced in Model II.
(See also the second column of the Table~\ref{tab:2}.)

In this context it is interesting to remark that
in nuclear matter the mass of the $\omega$-meson is reduced in 
Model~I but not in Model II. Thus, the nucleon is made of
heavier degrees of freedom in Model II. Heavier degrees of
freedom cause stronger internal forces \cite{Goeke:2007fq}, 
i.e.\ a larger magnitude of the pressure distribution which
we observe in Fig.~\ref{fig3}. (In principle, heavier degrees 
of freedom also contribute to a larger energy density in the
nucleon \cite{Goeke:2007fq}. However, in this work the
parameters are fixed to reproduce the same nucleon mass
in free space, and the same mass reduction for a nucleon bound  
in nuclear matter. Hence this effect on the energy density 
is not apparent.)

\begin{figure}
\includegraphics[scale=0.4]{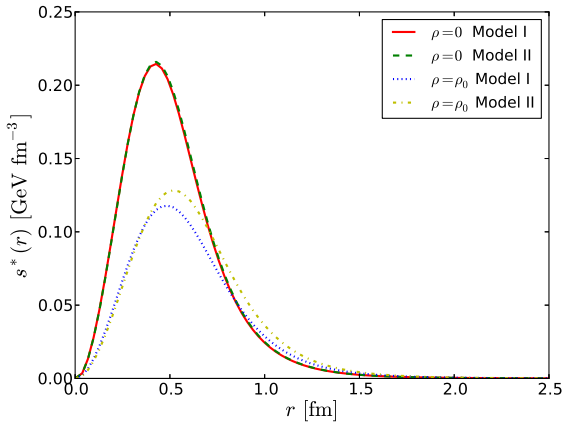}
\caption{The shear-force density $s^{*}(r)$ as a function of 
  radial distance $r$. The solid and dashed curves depict the
  densities respectively from Model I and Model II in free space. The
  dotted and dot-dashed ones represent respectively those  
  from Model I and Model II in nuclear matter. }
\label{fig4}
\end{figure}

In Fig.~\ref{fig4}, we draw the numerical results of the density of
shear forces. The conservation of the EMT implies
(in a static situation which we encounter here) 
that the spatial components of the EMT satisfy $\partial_i T_{ij}=0$.
Starting from Eq.~(\ref{Eq:30}), one then obtains the differential
equation   
\begin{equation}
  \label{eq:shear}
\frac23 \frac{\partial s^*(r)}{\partial r} + 2\;\frac{s^*(r)}{r} + 
\frac{\partial p^*(r)}{\partial r} \;=\; 0\,.
\end{equation}
It is straightforward to verify Eq.~(\ref{eq:shear}) by using 
the expressions (\ref{Eq:32},~\ref{Eq:33}) for $s^*(r)$ and $p^*(r)$ 
and the differential equations (\ref{eq:F}--\ref{eq:w}) for the 
profile functions $F(r)$, $G(r)$ and $\omega(r)$. Eq.~(\ref{eq:shear}) 
shows that  $s^*(r)$ and $p^*(r)$ are related to each other. 
However, the shear forces offer a new perspective and therefore 
equally instructive insights on the internal structure.
In order to discuss what we learn from $s^*(r)$, let us review the
liquid drop analogy. In a liquid drop $s^*(r)$ would be given by a
delta-function concentrated around the edge of the drop, and the
coefficient of the delta-function would be the surface tension. 
The liquid drop model was explored in \cite{Polyakov:2002yz}
to compute the $D$-terms of nuclei. Figure~\ref{fig4} shows that the
nucleon is much more diffuse than nuclei: 
the ``delta-function'' is strongly smeared out.
Nevertheless, the maximum of the shear-force density indicates the
``edge of the nucleon'' \cite{Goeke:2007fp}.
The effect of the nuclear medium is to shift the peak of $s^*(r)$ 
towards larger $r$ and broaden the shear force distribution.
So the nucleon swells and becomes even more diffuse in nuclear medium,
which confirms independently what we observed from other densities.
Notice that the peak of $s^*(r)$ is larger in Model II than in Model I,
as the higher ``surface tension'' has to oppose stronger inner
forces in Model II.
This reflects independently the stronger response of the 
nucleon to the stiffer nuclear environment in Model II.

\begin{figure*}
\includegraphics[scale=0.4]{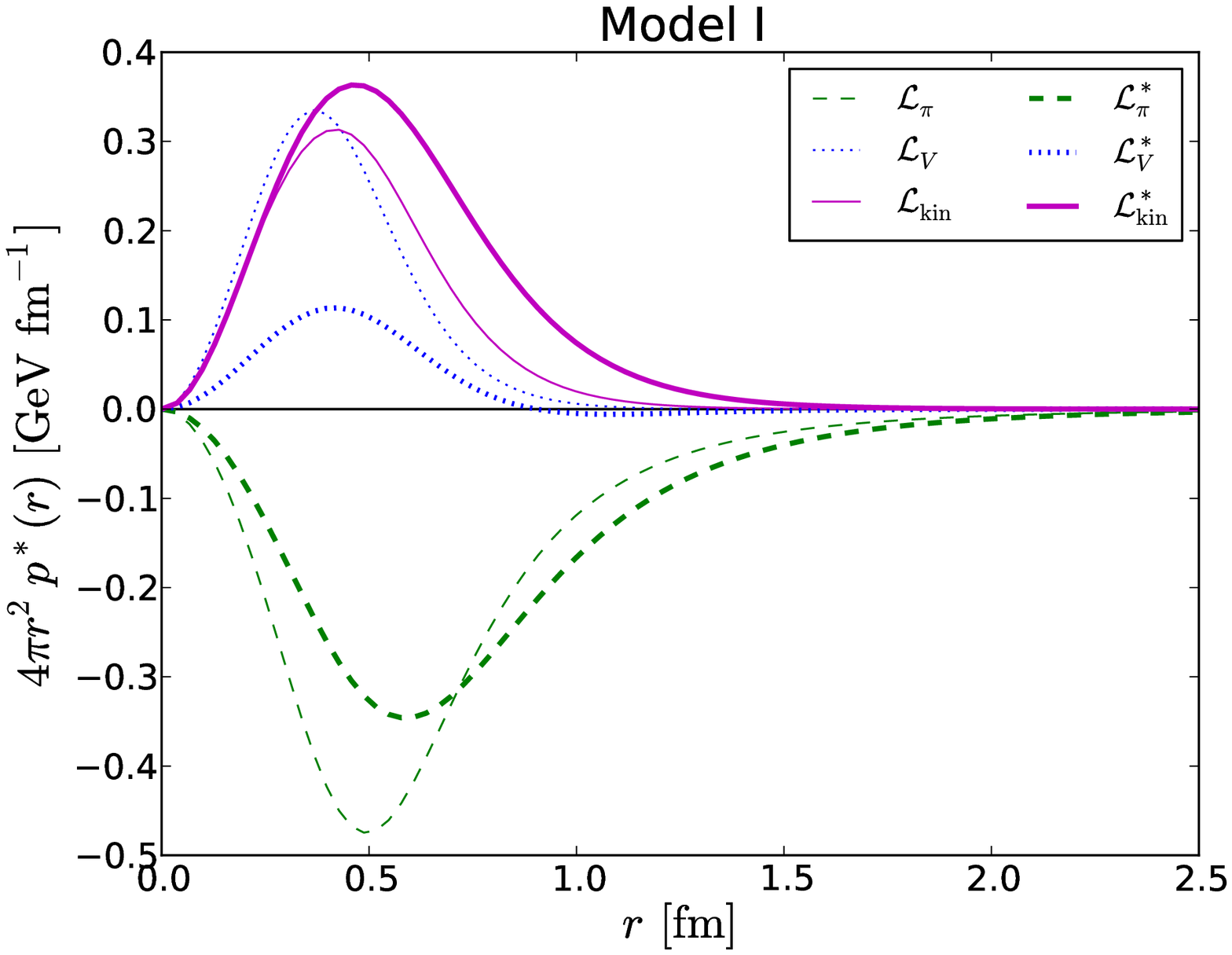} \hskip 1cm
\includegraphics[scale=0.4]{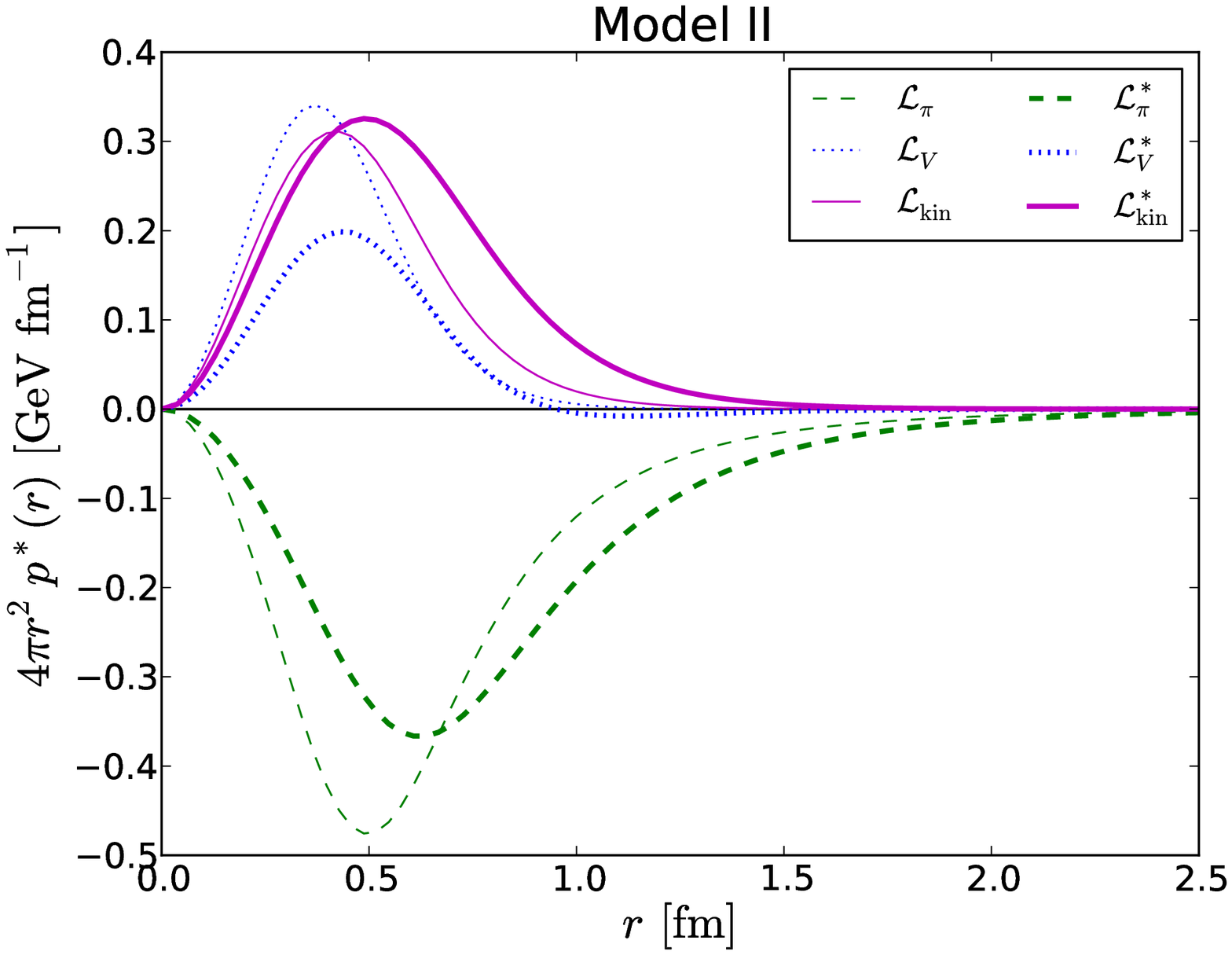} 
\caption{The contributions of each meson and the kinetic terms
to $4\pi r^2p^*(r)$. 
The left panel depicts the results from Model I and the right panel
does those from Model II. The dashed curve designates the pion 
contribution, the dotted one that of the vector mesons, and the 
solid one that of the kinetic terms. In both panels, the 
thin curves indicate the results in free space, while thick ones
designate those at normal nuclear matter density $\rho_0$. The total
results for $4\pi r^2p^*(r)$ were shown in Fig.~\ref{fig3}. }
\label{fig5}
\end{figure*}

In order to examine the modification of the nucleon in
detail, it is instructive to consider the contribution of each meson
to the pressure density.  In Fig.~\ref{fig5} we plot each contribution
of the pure pionic part ${\cal L}_\pi^*$, the pure vector meson part
${\cal L}_{\rm kin}^*$ and the interaction part ${\cal L}_V^*$ to the
pressure density, which discloses the most prominent feature of the
pressure density. As in free space, the pion furnishes a strong
attraction with the long-range asymptotics, which reflects the fact
that the soliton can never be stabilized by the pion only. Both the
vector mesons come into play to make the soliton
stable. On the other hand, the Skyrme quartic term stabilizes the
soliton in the Skyrme model. Thus, in the present model, the $\rho$
and $\omega$ mesons do play the same role as the Skyrme quartic  
term~\cite{Fujiwara:1984,Igarashi:1985,Meissner:1986ka}. The present  
results for the pressure distribution illustrate this well-known fact.  

The roles of the pion and the vector mesons were already discussed in
free space~\cite{Jung:2013bya}. Compared with the results in free
space
(indicated by the thin lines in Fig.~\ref{fig5}), in the nuclear 
medium the effect of the pion turns out to be noticeably suppressed, 
while the vector meson kinematic contribution is amplified, and the 
part due to the interaction term in turn is significantly suppressed. 
All contributions suffer a shift towards larger $r$, which illustrates 
the microscopic dynamics responsible for the swelling of the nucleon 
size in medium.

It is instructive to examine how the interaction part $\mathcal{L}_V^*$ 
is modified in nuclear medium. The contribution of $\mathcal{L}_V^*$ 
in nuclear matter is reduced and turns even negative before $r$ reaches
$1\,\mathrm{fm}$, see  Fig.~\ref{fig5}. This change is less pronounced
in Model~II than in Model~I, which reflects the fact that the $\omega$ 
meson mass is not modified in nuclear medium in Model~II. 
Therefore in Model II the core part of the in-medium nucleon 
is less modified compared to the free space case, 
see Fig.~\ref{fig5}.

Another consequence of keeping the omega parameters at their 
values in free space, is that the positive part of the pressure
density from $\mathcal{L}_{\rm kin}^*+\mathcal{L}_V^*$ turns out to be
larger in Model~II than in Model~I. 
This implies relatively larger repelling forces in the inner region,
and causes a larger swelling of the nucleon in 
Model~II as compared to Model~I.
(Compare the changes of all mean-squared radii in Model~I
with those in Model~II as listed in Table~\ref{tab:2}.)

\begin{figure}[b!]
\includegraphics[scale=0.4]{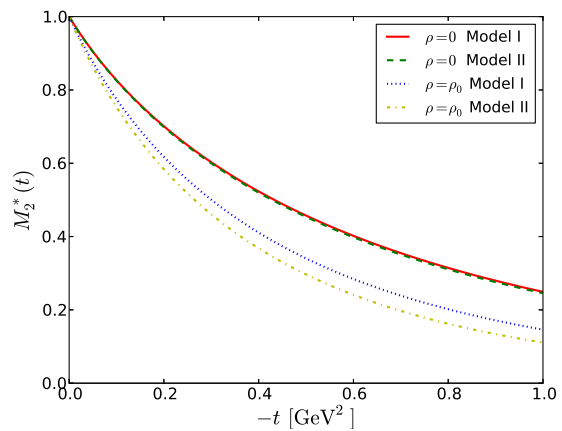}
\caption{$M_{2}^{*}(t)$ as a function of $t$. The solid
  and dashed curves depict the form factor respectively from Model I and
  Model II in free space. The dotted and dot-dashed ones represent
  respectively those from Model I and Model II in nuclear matter.
\label{M2}}
\end{figure}
\begin{figure}[b!]
\includegraphics[scale=0.4]{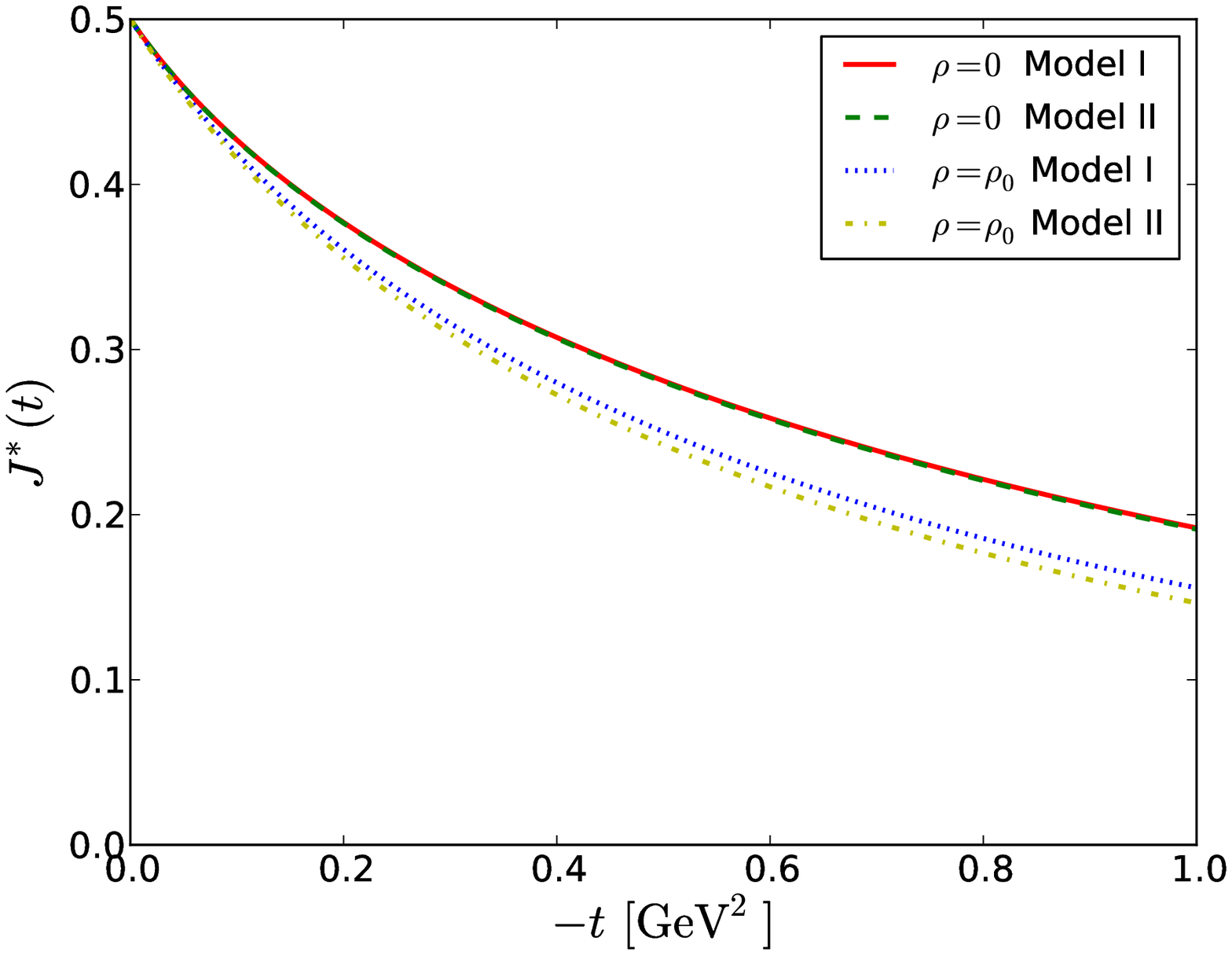}
\caption{ $J^{*}(t)$ as a function of $t$. The solid
  and dashed curves depict the form factor respectively from Model I and
  Model II in free space. The dotted and dot-dashed ones represent
  respectively those from Model I and Model II in nuclear matter.
\label{J}}
\end{figure}
\begin{figure}
\includegraphics[scale=0.4]{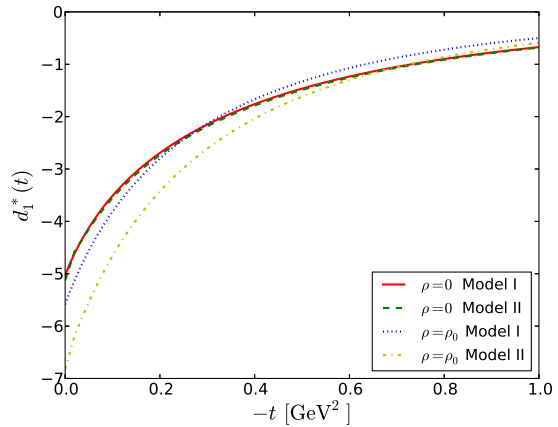}
\caption{$d_{1}^{*}(t)$ as a function of $t$. The solid
  and dashed curves depict the form factor respectively from Model I and
  Model II in free space. The dotted and dot-dashed ones represent
  respectively those from Model I and Model II in nuclear matter.
\label{d1}}
\end{figure}
With the densities discussed above, we can immediately derive the 
corresponding EMTFFs. In Figs.~\ref{M2}-\ref{d1} we depict the results
of the three EMTFFs defined in Eq.~(\ref{Eq:EMTff}). Note that the
$M_2^*(t)$ and $J^*(t)$ are constrained to be $1$ and $1/2$ at $t = 0$
respectively as shown in Eqs.~(\ref{eq:mass_con}) and
(\ref{eq:norm}). 
Similarly to the in-medium Skyrme model \cite{Kim:2012ts},
the mass form factor $M_2^*(t)$ and the spin form factor $J^*(t)$ 
in nuclear medium fall off more rapidly than those in free space. 
This reflects that the corresponding mean-squared radii in
nuclear matter become larger, as we have shown already in
Table~\ref{tab:2}.  

The last form factor, which is called the $D$-term form factor, is
distinguished from the other two form factors. It is not constrained
by any condition. However, the negative sign of $d_1$ at $t=0$, which
is called the $D$ term, arises as a consequence of stability. The $D$
term is defined in terms of the pressure density  
\begin{equation}
d_1^* \;=\; 5\pi M_N^* \int_0^\infty dr\,r^4  p^* (r).   
\label{eq:d1}
\end{equation}
Recall that the analog integral over $r^2  p^* (r)$ vanishes due
to the stability condition (\ref{Eq:39}).
This implies that $d_1^*$ must have a negative value, because $r^4$ in
the integrand of Eq.~(\ref{eq:d1}) lessens the contribution from the
inner (positive) part of the pressure density but amplifies 
that from the outer (negative) part. 
The explicit value of $d_1$ and its change  
in nuclear matter are given in Tables~\ref{tab:1} and~\ref{tab:2}. 
The magnitude of $d_1^*$ in nuclear matter is larger
than that in free space.  This can be understood from
Fig.~\ref{fig3} in which the pressure density in nuclear matter was
shown to be shifted to the outer region and became more strongly
negative as $r$ increases. As shown in Fig.~\ref{d1}, the
$d_1^*$ form factor in nuclear matter falls off more rapidly than that
in free space, as the other two form factors do.  

We note that the comparison of the results from Models I and II
explicitly shows that $d_1^*$ form factor is very sensitive to the
stabilization mechanism, i.e. to the interplay between the internal
and external forces which stabilize the soliton. This is seen from 
Table~\ref{tab:2}, where the change of $d_1^\ast$ in Model II is much
larger than in Model~I.  

Finally, we discuss the mean-squared radius of the EMT trace operator 
\cite{Goeke:2007fp}.
The form factor of the EMT trace operator can be expressed in terms 
of the other EMTFFs, so that the mean-squared radius 
$\langle r_F^2\rangle$ (in the notation of Ref.~\cite{Goeke:2007fp}) 
can be given as
\begin{equation}
    \langle r_F^2\rangle^\ast = \langle r_{00}^2\rangle^\ast 
    - \frac{12 d_1^\ast}{5M_N^{\ast 2}}.
\label{rF}
\end{equation}
The importance of this quantity lies in the fact that in QCD 
in the chiral limit the trace anomaly \cite{Adler:1976zt} relates 
it to the mean-squared radius of the gluonic operator
$G^{\mu\nu}G_{\mu\nu}$. Although explicit gluonic degrees of freedom
are absent, chiral soliton models allow one to evaluate consistently
the trace of the EMT, and thus to obtain in this way insights on this
quantity~\cite{Goeke:2007fp}. In Table~\ref{tab:1} we see that in
Model I and Model II $\langle r_F^2\rangle\,=\,1\,{\rm
  fm}^2$ in free space. A similar value was obtained in the Skyrme
model of Ref.~\cite{Kim:2012ts} (though it was not explicitly reported
there). In the chiral quark-soliton
model~\cite{Goeke:2007fp,Goeke:2007fq} and a different Skyrme
model study~\cite{Cebulla:2007ei}, comparable values were obtained for 
$\langle r_F^2\rangle$. It is an important observation that the
$\pi$-$\rho$-$\omega$ model confirms the results from other solitonic
approaches concerning the magnitude of $\langle r_F^2\rangle$. Thus,
the large value of the $\langle r_F^2\rangle$ appears to be a robust
prediction of chiral soliton models. Their prediction
for the mean-squared radius of the trace of the EMT operator is remarkable
\cite{Goeke:2007fp}, especially if one confronts it with the
QCD sum rule study in which the mean-squared radius of the 
{\it traceless} part of the gluonic contribution to the EMT
was predicted to be one-order-of-magnitude smaller
\cite{Braun:1992jp}.\footnote{ 
  The instanton vacuum approach provides a natural explanation
  for this observation. The trace of the EMT operator receives 
  one-instanton contributions. In contrast, the traceless part 
  arises from instanton anti-instanton configurations
  \cite{Goeke:2007fp} and is therefore of higher order in the 
  instanton packing fraction, which is the small parameter in 
  the instanton vacuum approach describing the diluteness
  of the instanton medium \cite{Diakonov:1995qy}.}

In our context, it is also interesting to explore the effects of the
nuclear environment on the trace of the EMT operator.
Table \ref{tab:2} shows that nuclear medium effects increase the
mean-squared radius of the EMT trace operator by 30--40$\,\%$ in 
Model I and II in the $\pi$-$\rho$-$\omega$-soliton framework,
and similarly in the Skyrme model \cite{Cebulla:2007ei}. 
This is rather remarkable. At this point one has to be cautious, 
because one should take the chiral limit in order to isolate a gluonic  
contribution to the EMT trace operator. This will be an interesting
task for future studies. Here we content ourselves to
remark that, by exploring the trace anomaly, studies of the EMT
provide a unique opportunity to learn about nuclear  
medium effects on certain gluonic quantities. 

\section{Conclusions and Summary}
The present work aimed at investigating the energy-momentum tensor
form factors in nuclear matter within the framework of the 
medium-modified $\pi$-$\rho$-$\omega$ soliton model. The parameters 
in the model have been fixed by using the experimental data. The medium
functionals, which are introduced to describe the influence of a nuclear
environment on the properties of the single soliton, have been fixed
by using the analysis of pionic atoms and the data on low-energy
pion-nucleus scattering. We set up two different models,
Model I and Model II,
by making different assumptions on the KSRF relation.  
Both provided comparably satisfactory descriptions of medium
effects, where Model II describes a stiffer nuclear medium.

We computed the densities associated with the energy momentum tensor,
and found that they all exhibited qualitatively similar patterns in
nuclear medium: they are diminished in the inner region of the nucleon
and broadened towards larger distances. This nicely illustrates the
well-known facts that the nucleon swells and its mass is reduced in
nuclear matter. 
In particular, the discussion of the pressure distribution has shown 
in great detail how the nucleon is stabilized in nuclear medium by 
the interplay of the different degrees of freedom within the 
$\pi$-$\rho$-$\omega$ soliton model. 
While the description of the energy-momentum tensor densities is
qualitatively similar in Model~I and Model~II, we made the
interesting observation that a stiffer nuclear medium causes the nucleon to
experience stronger internal forces.
We also derived the three energy-momentum tensor form factors
$M_2(t)$, $J(t)$, $d_1(t)$, and showed that they fall off faster than
those in free space, which reflects that in general the mean-squared
radii become larger in nuclear medium.

In all theoretical studies so far, the $D$-term 
$d_1\equiv d_1(0)$ was found to be negative, for the 
nucleon \cite{Ji:1997gm,Ossmann:2004bp,Wakamatsu:2006dy,
Goeke:2007fp,Goeke:2007fq,Cebulla:2007ei,Jung:2013bya},
the pion~\cite{Polyakov:1999gs}, nuclei~\cite{Polyakov:2002yz,Guzey:2005ba},
photons~\cite{Gabdrakhmanov:2012aa}, and $Q$-balls~\cite{Mai:2012yc}.   
The negative sign of the $D$-term was confirmed also for a nucleon
bound in nuclear medium in Ref.~\cite{Kim:2012ts} and in the 
present work. Medium effects do not change the generic pattern
how internal forces balance to form a stable nucleon.
However, they alter the strengths and ranges of the various
contributions from the $\pi$-, $\rho$- and $\omega$-meson degrees of
freedom. The $D$-term is the quantity most sensitive to modifications 
in the interplay of attractive and repulsive internal forces
inside the nucleon. The results obtained in this work and in
the medium-modified Skyrme model~\cite{Kim:2012ts} will shed light on 
nuclear medium effects on the energy-momentum tensor 
form factors.

To summarize, 
the study of the energy-momentum tensor form factors reveals 
directly the internal structure of the nucleon in both free space and
in nuclear matter, of which the latter was in the focus of the present
work. It is of great importance to investigate the nucleon inside the
isospin asymmetric matter~\cite{Yakhshiev:2013eya}, 
in neutron stars as well as inside real nuclei, where 
surface effects become essential. Related works are under way.  

\section*{Acknowledgments}

H.-Ch. K is grateful to Kyungseon Joo for his hospitality during his
visit to University of Connecticut and P. Navratil and R. Woloshyn for
their hospitality during his visit to TRIUMF, where parts of the work
have been done. This work is supported by the Basic Science Research
Program through the National Research Foundation (NRF) of Korea funded
by the Korean government (Ministry of Education, Science and
Technology), Grant No.~2012-0008469 (J.H.J. and U.Y.) and Grant
No.~2012004024 (H.Ch.K.). The work was partly supported by DOE
contract DE-AC05-06OR23177, under which Jefferson Science Associates,
LLC, operates the Jefferson Lab.

\

\appendix
\section{Proof of von Laue condition}
\label{App:A}

In this Appendix we present an alternative proof of the
von Laue condition in Eq.~(\ref{Eq:39}).
We substitute $r\to r^\prime =\lambda r$ in Eq.~(\ref{Eq:XXXX}) 
(and drop the prime on the integration variable
$r^\prime$ for simplicity). This yields
\begin{widetext}
\begin{eqnarray}
\label{Eq:App-1}
M_{\mathrm{sol}}^*(\lambda) &=&
\lambda\;\,
4\pi \int_0^\infty {\rm d}r\, r^2\,
\left\{
\frac{1}{g_\rho^{2}\zeta_\rho
  r^{2}}\left(G'^{2}+\frac{G^{2}\left(G+2\right)^{2}}{2r^2}\right)\right\} 
\cr
&+& 
\lambda^0
4\pi \int_0^\infty {\rm d}r\, r^2\,
\left\{
\left(\frac{3}{2}g_\omega\sqrt{\zeta_\omega}\right)\frac{1}{2\pi^{2}r^{2}}
\omega\sin^{2}F\, F'\right\}\cr
&+& 
\frac{1}\lambda\;4\pi\int_0^\infty {\rm d}r\, r^2\,
\left\{
    \alpha_{p}f_{\pi}^{2}\left(\frac{\sin^{2}F}{r^{2}}+\frac{F'^{2}}{2}\right)
    +\frac{2f_{\pi}^{2}}{r^{2}}\left(1-\cos F+G\right)^{2}-\frac{1}{2}\omega'^{2}
\right\}
\cr
&+&\frac{1}{\lambda^3} \,4\pi\!\int_0^\infty {\rm d}r\, r^2\,
\biggl\{
    \alpha_{s}f_{\pi}^{2}m_{\pi}^{2}\left(1-\cos F\right)
    -\zeta_\omega g_\omega^{2}f_{\pi}^{2}\omega^{2}
\biggr\} \,.
\end{eqnarray}
The transformation $r\to \lambda r$ corresponds to a dilatational variation.
The soliton solution is a minimum of the energy functional for any type of
variations. Therefore, $M_{\mathrm{sol}}^*(\lambda)$ must have a minimum at 
$\lambda=1$, which implies that
\begin{eqnarray}
\label{Eq:App-2}
\frac{{\rm d}M_{\mathrm{sol}}^*(\lambda)}{{\rm d}\lambda}\biggl|_{\lambda=1} &=&
4\pi \int_0^\infty {\rm d}r\, r^2\,
\left\{
\frac{1}{g_\rho^{2}\zeta_\rho
  r^{2}}\left(G'^{2}+\frac{G^{2}\left(G+2\right)^{2}}{2r^2}\right)\right\} 
\cr
&-& 
4\pi\int_0^\infty {\rm d}r\, r^2\,
\left\{
    \alpha_{p}f_{\pi}^{2}\left(\frac{\sin^{2}F}{r^{2}}+\frac{F'^{2}}{2}\right)
    +\frac{2f_{\pi}^{2}}{r^{2}}\left(1-\cos F+G\right)^{2}-\frac{1}{2}\omega'^{2}
\right\}
\cr
&-&3\times4\pi\!\int_0^\infty {\rm d}r\, r^2\,
\biggl\{
    \alpha_{s}f_{\pi}^{2}m_{\pi}^{2}\left(1-\cos F\right)
    -\zeta_\omega g_\omega^{2}f_{\pi}^{2}\omega^{2}
\biggr\} \nonumber\\
&=& 3\times4\pi\!\int_0^\infty {\rm d}r\, r^2\,p^{*}(r) \nonumber\\
&\stackrel{!}{=}& 0 \, .
\end{eqnarray}
\end{widetext}
In the last step, we explored the fact that the integrand in the
energy functional is the same as the expression for the pressure
density in Eq.~(\ref{Eq:33}) with factor 3 multiplied. 
This proves the von Laue condition in Eq.~(\ref{Eq:39}). 

The dilatational variation is a standard procedure to prove the
existence  
of an energy extremum. Equation~(\ref{Eq:App-2}) implies a relation
among the different contributions to the soliton energy, which is
sometimes known as a ``virial theorem''.  Analog proofs of the von
Laue condition were formulated in the chiral quark-soliton model
\cite{Goeke:2007fp} and Skyrme model \cite{Cebulla:2007ei}.  

Notice that the Wess-Zumino term does not contribute to $p^*(r)$ 
because the omega-field $\omega(r) \delta_{\mu0}$ has
no spatial components in the Ansatz (\ref{Eq:14}) and hence makes
no contribution to the stress tensor. In our proof, we see that the 
Wess-Zumino term drops out from the von Laue condition (\ref{Eq:39})
since its contribution to the soliton energy scales as $\lambda^0$ in
$M_{\mathrm{sol}}^*(\lambda)$ and therefore does not appear in the
dilatational variation in Eq.~(\ref{Eq:App-2}).

\

\end{document}